\documentclass[11pt]{article}

\usepackage{bbm,epic,eepic,framed,epsf,latexsym,amsmath,amssymb,amscd,mathrsfs,slashbox }
\usepackage{graphicx}
\usepackage[pdftex,colorlinks]{hyperref}

\usepackage[margin=1.0in]{geometry}
\newtheorem{theorem}{Theorem}[section]
\newtheorem{lemma}[theorem]{Lemma}

\newtheorem{note}[theorem]{Note}

\newtheorem{counter-example}[theorem]{Counter example}
\newtheorem{proposition}[theorem]{Proposition}
\newtheorem{open question}[theorem]{Open question}
\newtheorem{corollary}[theorem]{Corollary}

\newtheorem{definition}[theorem]{Definition}

\newtheorem{observation}{Observation}

\newcommand{\ignore}[1]{}

\newcommand{\cd}{{\cal D}}

\newcommand{\cp}{{\cal P}}

\newcommand{\reals}{{\mathbb R}}

\newcommand{\proof}{{\par\noindent {\bf Proof}\space\space}}
\newcommand{\proofbox}{\begin{flushright}$\Box$\end{flushright}}

\title{On the practically interesting instances of MAXCUT}

\begin{document}

\author{Yonatan Bilu \thanks{Parasight inc,
Agudat sport hapoel 1, Jerusalem, Israel.  yonatan@gmail.com}
                 \and Amit Daniely \thanks{Department of Mathematics, Hebrew
                 University, Jerusalem 91904, Israel. Supported in part by a binational Israel-USA grant 2008368. amit.daniely@math.huji.ac.il}
                 \and Nati Linial \thanks{School of Computer Science and Engineering, Hebrew University, Jerusalem 91904, Israel. Supported in part by a binational Israel-USA grant 2008368.  nati@cs.huji.ac.il}
                 \and Michael Saks \thanks{Department of Mathematics, Rutgers University, Piscataway, NJ 08854.  Supported in part by NSF under
grant  CCF-0832787 and by a binational Israel-USA grant 2008368. saks@math.rutgers.edu.}}

\maketitle

\setcounter{page}{0}

\thispagestyle{empty}
\maketitle

\begin{abstract}
The complexity of a computational problem is traditionally quantified based on the hardness of its worst case. This approach has many advantages and has led to a deep and beautiful theory. However, from the practical perspective, this leaves much to be desired. In application areas, practically interesting instances very often occupy just a tiny part of an algorithm's space of instances, and the vast majority of instances are simply irrelevant. Addressing these issues is a major challenge for theoretical computer science which may make theory more relevant to the practice of computer science.

Following \cite{BL}, we apply this perspective to MAXCUT, viewed as a clustering problem. Using a variety of techniques, we investigate practically interesting instances of this problem. Specifically, we show how to solve in polynomial time distinguished, metric, expanding and dense  instances of MAXCUT under mild stability assumptions. In particular, $(1+\epsilon)$-stability (which is optimal) suffices for metric and dense MAXCUT. We also show how to solve in polynomial time $\Omega(\sqrt{n})$-stable instances of MAXCUT, substantially improving the best previously known result.
\end{abstract}

\newpage

\section{Introduction}

The primary criterion used in computational complexity to evaluate algorithms is worst case
behavior, so that a problem is infeasible if
no efficient algorithm can solve {\em all} its instances. In practice, this approach is often overly pessimistic, and
a more realistic (but fuzzy) criterion would be to say that a problem is feasible if 
there is an efficient algorithm that correctly solves all of its {\em practically interesting} instances. 
The difference can be very substantial, since for many computational problems, the vast majority of 
instances are completely irrelevant for practical purposes.

An important case in point is clustering, where one seeks a meaningful partition of a given set of data. 
Almost every formal manifestation of the clustering problem is $NP$-Hard, yet, a clustering instance  
is of practical interest only if the data {\em can indeed} be partitioned in a meaningful way. Random 
instances are not likely to have a meaningful partition, so data sets with a meaningful partition are very special. 
Thus, even if no efficient algorithm can find the optimal partition for {\em every} data set, 
this does not imply that clustering is hard in practice. As Tali Tishby put it in conversation many years ago, 
many practitioners hold the opinion that "clustering is either easy or pointless". That is, for a data sets that 
admit a meaningful partition of the data, finding it is not hard.

Can this intuition be put on  a solid theoretical foundation? 
Bilu and Linial \cite{BL} proposed a framework for studying this issue. Generally speaking, their approach pertains to optimization 
problems with a continuous input space and discrete solution
space. They proposed two criteria for an optimal solution to be {\em evidently} optimal. A solution is {\em stable} if it remains optimal under moderate perturbations of the input. A solution is {\em distinguished} if a transition to another solution reduces the value of the objective function in proportion to the distance between the two solutions. Concretely, they considered the case where the input is a weighted graph and the candidate solutions are cuts (or more generally, partitions). Here, a cut is $\gamma$-stable (for $\gamma\ge 1$) if it remains optimal even if each input weight $w_{ij}$ is
perturbed to a value between $w_{ij}$ and $\gamma w_{ij}$. A cut is $\alpha$-{\em distinguished}  (for $\alpha\ge 0$) if moving to any other cut reduces the objective function by at least $\alpha$ times the sum of (weighted) degrees of the vertices that switched side.

Following Bilu and Linial \cite{BL}, we apply these notions to the study of the (weighted) MAXCUT problem. We also investigate the more restricted problem of Metric-MAXCUT\footnote{That is, MAXCUT, restricted to instances where the weight function is a metric.} which arises often in the field of machine learning. 
Our main results are:

\begin{theorem}\label{thm:main_metric}
\begin{enumerate}
\item For every $\epsilon>0$ there is a polynomial time algorithm that correctly solves all $(1+\epsilon)$-locally stable instances of Metric-MAXCUT.
\item For every $\epsilon>0$ and $C>1$ there is a polynomial time algorithm that correctly solves all $(1+\epsilon)$-locally stable and $C$-dense instances of MAXCUT.
\end{enumerate}
\end{theorem}
The condition of $C$-density rules out overly-weighted edges. The notion of $\gamma$-local stability is a substantial weakening of $\gamma$-stability. It is defined similarly, but we only require resilience to perturbations that modify edges which are all incident with the same vertex.

\begin{theorem}\label{thm:main_distinct}
There is a polynomial time algorithm that solves all instances of MAXCUT that are
\begin{enumerate}
\item $\alpha$-distinguished and $\gamma$-locally stable with $\gamma >  \frac{2}{1-\sqrt{1-\alpha^2}}$, or
\item $\gamma$-locally stable with $\gamma > \frac{2}{1-\sqrt{1-h^2}}$.
\end{enumerate}
Here $h$ is the Cheeger constant of the maximal cut.
\end{theorem}
This substantially improves a result from \cite{BL} that works only for regular graphs  and requires that $\gamma>\frac{5+\sqrt{1-\alpha^2}}{1-\sqrt{1-\alpha^2}}$ or $\gamma>\frac{5+\sqrt{1-h^2}}{1-\sqrt{1-h^2}}$. It is also shown in \cite{BL} that $n$-stable instances are feasible. Here we derive the same conclusion under the weaker (but still impractical) assumption of $\Omega(\sqrt{n})$-stability.
\begin{theorem}\label{thm:main_stable}
There is a polynomial time algorithm that finds the optimal solution for every $\Omega(\sqrt{n})$-stable instance of MAXCUT.
\end{theorem}

\subsubsection*{Some notation and terminology}
Here the input to the MAXCUT problem is the complete graph on $n$ vertices
$G=(V,E)$ along with a symmetric function  with zero diagonal $w:V\times V\to\mathbb{R}^+$. Expressions such as "$w$ is bipartite" refer to the graph which is the support of $w$, which is always assumed to be connected. Our purpose is to find a {\em cut} $(S,\bar S),\; S\subseteq V$ for which $\sum_{a \in S,~ b  \in \bar S} w(a,b)$ is maximized.

Fix a cut $(S, \bar S)$. We use the self-explanatory terms ``the vertices $x, y$ are {\em on the same side}" or {\em ``separated"} by this cut. We call the edge $xy$ a {\em cut edge} or a {\em non-cut edge} when $x, y$ are separated resp. on the same side of the cut. For $A,B\subset V$, we denote $E(A,B)=\{ab|a\in A,~ b\in B\}$ and $w(A,B):=\sum_{uv \in E(A,B)}w(u,v)$. Also $\tau_w(A)=\tau(A)=w(A,\bar A)$ and $\mu(A)=\mu_w(A)=w(A,V)$.
Let $A \subseteq V$. We denote by $\xi (A)$, the weight of the cut edges emanating from $S$, i.e., $\xi(A)=\sum_{vu\in E(A,\bar A)\cap E(S, \bar S)}w(u,v)$ and by $\iota(A)=\tau(A)-\xi(A)$ the weight of the non-cut edges. We slightly abuse notation for singletons $A=\{v\}$ and pairs $A=\{u, v\}$ and write $\tau(v)$ or $\iota(e)$ etc., where $e=uv$. The {\em minimal, maximal} and {\em average} degree of $w$ are denoted by $\underline{\delta}(w)=\min_{v\in V}\mu(v)$, $\bar{\delta}(w)=\max_{v\in V}\mu(v)$ and $\delta(w)=\frac{\sum_{v\in V}\mu(v)}{n}$ respectively. (The potentially confused reader may find the following Greek-mathematical dictionary useful: $\tau$ stands for ``total", $\xi$ for ``external" and $\iota$ for ``internal").

\subsection{Stable instances}
\begin{definition}
Let $w:V\times V\to [0,\infty)$ be an instance of MAXCUT and let $\gamma \ge 1$. An instance $w':V\times V\to [0,\infty)$ is a {\em $\gamma$-perturbation} of $w$ if
$$\forall u,v\in V,\;w(u,v)\le w'(u,v)\le \gamma\cdot w(u,v)$$
An instance $w$ is said to be {\em $\gamma$-stable} if there is a cut which forms a maximal cut for every $\gamma$-perturbation $w'$ of $w$.
\end{definition}
\begin{definition}
Let $\gamma \ge 1$. An instance $w:V\times V\to [0,\infty)$ for MAXCUT  is {\em $\gamma$-locally stable} if there is a maximal cut $(S,\bar S)$ for which it is impossible to obtain a larger cut by switching the side of some vertex $x$ and multiplying the edges in $E(x,V\setminus \{x\})$ by numbers between $1$ and $\gamma$.
\end{definition}
The definitions of stability and local stability capture the intuition of an ``evidently optimal" solution. The following more concrete equivalent definitions are usually more convenient to use.
\begin{observation}\label{observation:equiv_def}
\cite{BL} Let $w:V\times V\to \mathbb R$ be an instance of MAXCUT and let $\gamma\ge 1$.
\begin{itemize}
\item The instance $w$ is $\gamma$-stable iff there is a maximal cut for which $\xi(A)\ge \gamma\cdot \iota(A)$ for every $A\subset V$.
\item The instance $w$ is $\gamma$-locally stable iff there is a maximal cut for which $\xi(x)\ge \gamma\cdot \iota(x)$ for every $x\in V$.
\end{itemize}
\end{observation}
We say that a (not necessarily maximal) cut $(S,\bar S)$ is {\em $\gamma$-stable} (resp. {\em $\gamma$-locally stable}) if the first (resp. second) condition in Observation \ref{observation:equiv_def} holds.

As Observation \ref{observation:equiv_def} shows, every instance is $1$-stable, and being $\gamma$-stable for {\em some} $\gamma>1$ is equivalent to having a unique maximal cut\footnote{To see that, note that if $(S,\bar S)$ is a $\gamma$-stable cut and $(T,\bar T)$ is another cut then $w(T,\bar T)=w(S,\bar S)-\xi((T\cap S)\cup(\bar T\cap \bar S))+\iota((T\cap S)\cup(\bar T\cap \bar S))\le
w(S,\bar S)-\frac{\gamma-1}{\gamma+1}\tau((T\cap S)\cup(\bar T\cap \bar S))<w(S,\bar S)$.}. Finally, an instance is bipartite iff it is $\gamma$-stable for {\em every} $\gamma\ge 1$. Thus, $\gamma$-stability is seen to be a relaxation of being bipartite.

Stability and local stability are quite different. As mentioned, for $\gamma > 1$ every instance has at most one $\gamma$-stable cut. On the other hand, there can be numerous $\gamma$-locally stable cuts: Consider the instance where $w=1$ on the edges of a perfect matching and $\epsilon > 0$ elsewhere. As $\epsilon \to 0$, the local stability tends to $\infty$. Yet, this instance is not $\gamma$-stable for any $\gamma>1$. It is easy to check that this instance has exponentially many $\gamma$-locally stable maximal cuts. From the computational perspective the two properties are very different as well. Thus MAXCUT remains $NP$-hard even under arbitrarily high local stability (see \cite{BL}), whereas we show here how to efficiently solve $\Omega(\sqrt{n})$-stable instances. Also, it is easy to decide whether a given cut is $\gamma$-locally stable, but we do not know how to decide whether a given cut is $\gamma$-stable and we suspect that this problem is hard.

In Section~\ref{sec:general}, following a simplified version of the algorithm in~\cite{BL} for $\Omega(n)$-stable instances, we present a deterministic algorithm that solves every $\Omega(\sqrt{n})$-stable instance, proving Theorem \ref{thm:main_stable}.

\subsection{Distinguished and Expanding instances}

Let $w:V\times V\to\mathbb R^+$ be an instance of MAXCUT whose (unique) maximal cut is $(S,\bar S)$. We note that if all vertices of $A\subset V$ switch side, then the weight of the cut decreases by $\xi(A)-\iota(A)$. Thus, we define
\begin{definition}
An instance $w$ of MAXCUT is {\em $\alpha$-distinguished} for $\alpha\ge 0$ if for every $\emptyset\ne A\subset V$, $ \xi(A)-\iota(A)\ge \alpha\cdot \min \{\mu(A),\mu(\bar A)\}$.
\end{definition}

Note that every instance is $0$-distinguished and being $\alpha$-distinguished with $\alpha>0$ is equivalent to having a unique maximal cut. It is not hard to see that $\frac{1+\alpha}{1-\alpha}$-local stability is equivalent to $\alpha$-{\it local distinction}, namely $\xi(x)-\iota(x)\ge \alpha\cdot \mu(x)$ for every $x\in V$.

{\bf Distinction vs Stability.}
Let $(S,\bar S)$ be a maximal cut of $w:V\times V\to [0,\infty)$. On the one hand, every $\alpha$-distinguished instance is $\frac{1+\alpha}{1-\alpha}$-stable, because $\xi(A)-\iota(A)\ge \alpha \mu(A)\ge \alpha(\xi(A)+\iota(A))$. On the other hand, highly stable instances need not be distinguished as the following bipartite example with $V=\{a_1,\ldots,a_n\}\dot\cup\{b_1,\ldots,b_n\}$ shows. Here $w(a_i,b_j)$ is $1$ when $i=j$ and $\epsilon \ll 1$ otherwise. Clearly $w$ is $\infty$-stable. Yet, switching the sides of all the vertices in $\{a_1,\ldots,a_{\frac{n}{2}}\}\cup\{b_1,\ldots,b_{\frac{n}{2}}\}$ decreases the weight of the cut only slightly. Such examples motivate the stronger notion of distinction. Although the cut $(\{a_1,\ldots,a_n\},\{b_1,\ldots,b_n\})$ is infinitely stable, its optimality does not seem completely evident.

{\bf Distinction and Expansion.} Call $w:V\times V\to \mathbb R^+$ {\em $\beta$-expanding} if $\beta\le h(w)$ where $h(w)=\min_{\emptyset\ne A\subset V}\frac{\tau(A)}{\min \{\mu(A),\mu(\bar A)\}}$ is $w$'s {\em Cheeger constant}. An $\alpha$-distinguished instance is $\alpha$-expanding, though highly expanding instances can even have multiple maximal cuts. However, an instance that is both $\gamma$-stable and $\beta$-expanding is easily seen to be $(\beta\cdot \frac{\gamma -1 }{\gamma+1})$-distinguished. As this discussion implies, distinction is a conjunction of stability and expansion.

In section \ref{sec:distinct} we prove Theorem \ref{thm:main_distinct}, using a spectral result from \cite{BL}. In the appendix we re-derive this result and point out its close relation to the Geomans-Williamson algorithm \cite{GM} and other spectral techniques.

\subsection{Metric and Dense instances}
In Section \ref{sec:metric_and_dense} we study metric instances. This is done through a reduction from metric to dense instances, so we consider such instances as well (Section \ref{sec:dense}).

We call $w:V\times V\to \mathbb R$ {\bf $C$-dense} for $C\ge 1$ if $\forall x,y\in V,\; w(x,y)\le C\cdot \frac{\tau(x)}{n}$.
As shown in \cite{AKK}, for $C>1$ fixed, $C$-dense MAXCUT is $NP$-Hard, but it has a PTAS. As we show, this PTAS can be adapted to correctly solve all instances of MAXCUT that are $(1+\epsilon)$-locally stable and $C$-dense for every $\epsilon >0, C>1$. The algorithm samples $O(\log n)$ vertices and tests each of their bipartitions as a seed to a cut. As we show, w.h.p., one of the resulting cuts is the maximal cut, proving the second part of Theorem \ref{thm:main_metric}.

In Section \ref{sec:metric} we deal with Metric-MAXCUT. As shown in \cite{VK} (with credit to L. Trevisan) Metric-MAXCUT is $NP$-Hard. That paper also gives a reduction from metric to $(4+o(1))$-dense instances of MAXCUT, thus yielding a PTAS for Metric-MAXCUT. We show that a slight variation of this reduction preserves local stability\footnote{A word of caution: Our definition of stability and local stability for Metric-MAXCUT is more restrictive than one might think. We require the perturbed instance to satisfy the stability condition {\em whether or not} it is metric.}, and therefore yields an efficient algorithms for $(1+\epsilon)$-locally stable instances of Metric-MAXCUT, proving Theorem \ref{thm:main_metric} in full.

This algorithm for metric instances is far from being a practically applicable clustering method. Even though it is polynomial-time, the actual run times are prohibitively high. We view this more as an invitation to seek practical algorithms for $\gamma$-stable instances of metric MAXCUT for some reasonable values of $\gamma$. Specifically we provide such an algorithm for $(3+\epsilon)$-locally stable metric instances.

\subsection{Relation with other work}\label{sec:other_work}
Smoothed analysis is the best known example of a method for analyzing instances of computational problems based on their practical significance. As this method shows~\cite{ST}, a certain variant of the simplex algorithm solves in polynomial time {\em almost every} input. Even closer to our theme are several recent papers on clustering. In \cite{ABS} polynomial time algorithms are given for $3$-stable instances of $k$-means, $k$-medians and other ``center based" clustering problems. The constant $3$ was improved in \cite{BL2} to $(1+\sqrt{2})$ for $k$-median. The papers \cite{DLS, AB, BBV} consider data sets that admit a good clustering and show how to cluster them efficiently.

Also related to our work are the {\em planted partition model} \cite{B} and {\em semirandom model \cite{FK}} for MAXCUT. In these models instances are generated by splitting the vertices at random $V=S \dot \cup \bar S$. Edges in $S\times\bar S$ (resp. $S\times S\cup \bar S\times \bar S$) are picked with probability $p$, resp. $q<p$. In the semirandom model we also allow an adversary to add edges to $S\times\bar S$ and drop edges from $S\times S\cup \bar S\times \bar S$. As shown in \cite{B,FK}, a.a.s., $(S,\bar S)$ is the maximal cut and it can be efficiently found using certain algorithms. It not hard to see that for fixed $p$ and $q$, this is a consequence of Theorem \ref{thm:main_metric}. 
The planted partition model is a random model that usually generates instances with a good partition, and those can be efficiently found. The semirandom model goes further by allowing an adversary to modify the input in a way that improves the optimal partition. Here we take an additional step forward, since we solve efficiently {\em every} instance with a good partition.

\section{Algorithms for locally stable dense and metric instances}\label{sec:metric_and_dense}

\subsection{Dense instances}\label{sec:dense}
\begin{theorem}\label{theorem:dense}
For every $C\ge 1$ and $\epsilon>0$
there is a randomized polynomial time algorithm that correctly solves all $(1+\epsilon)$-locally stable, $C$-dense instances of MAXCUT.
\end{theorem}

The analysis of the algorithm is based on the following lemma.
\begin{lemma}\label{lemma:sampling}
Suppose that $w:V\times V\to [0,\infty)$ is a $C$-dense instance and let $(S,\bar S)$ be a $\gamma$-locally stable cut. Let $X_1,\ldots,X_m$ be i.i.d. r.v. that are uniformly distributed on $V$. For $x\in V$,
let $A_x$ be the event that $S_{+} > S_{-}$, where $S_{\pm} = \sum  w(x,X_i)$ over all $i$ s.t. $x$ and $X_i$ are separated resp. on the same side. Then
$$\Pr\left( \cup_x A_x\right)\le |V|\cdot\exp \left(-\frac 12\left(\frac{1}{C}\cdot \frac{\gamma-1}{\gamma+1}\right)^2 \cdot m\right)$$
\end{lemma}
\proof
The lemma follows from  Hoeffding's bound. For every $x\in V$, $S_{+} - S_{-}$ is a sum of $m$ i.i.d. r.v.'s of expectation $\frac{\xi(x)-\iota(x)}{|V|}\ge \frac{\gamma -1}{\gamma+1}\frac{\tau(x)}{|V|}$. These r.v.'s are bounded in absolute value, by $C\cdot \frac{\tau(x)}{|V|}$.
\proofbox

\proof (Of Theorem \ref{theorem:dense})
Let $D=2 \left( C\cdot\frac{2+\epsilon}{\epsilon}\right)^2$. Let $m=D\cdot \ln (2|V|)$.
Take an i.i.d. sample of $m$ uniformly chosen points $X_1,\ldots,X_m\in V$. By the above lemma, with probability $\ge 0.5$, there is a partition $\{X_1,\ldots, X_m\}=L\coprod R$ such that the cut defined by
$S=\{x\in V: w(x,R)>w(x,L)\}$
is the optimal cut. Since the number of such partitions is  $(2\cdot |V|)^{\ln (2)D}$, there are only polynomially many partitions to consider, yielding an efficient randomized algorithm for the problem.
\proofbox

\begin{corollary}
For every $C\ge 1$ and $\epsilon>0$, a $C$-dense instance of MAXCUT has only $\mbox{poly}(|V|)$-many $(1+\epsilon)$-locally stable cuts.
\end{corollary}
\proof
Consider the random cut $(S,\bar S)$, sampled as in the proof of Theorem \ref{theorem:dense}, where the partition $\{X_1,\ldots, X_m\}=L\coprod R$ is chosen uniformly at random. The proof Theorem \ref{theorem:dense} shows that for every $(1+\epsilon)$-locally stable cut $(T,\bar T)$, the probability  that $(S,\bar S)=(T,\bar T)$ is $\ge 0.5\cdot (2\cdot |X|)^{-D\ln(2)}$. Thus, there are at most $2 \cdot (2\cdot |X|)^{D\ln(2)}$ such cuts.
\proofbox

\subsection{Metric instances}\label{sec:metric}
Given an instance $w:V\times V\to[0,\infty)$ of MAXCUT, we split its vertices as follows. Pick a set $\tilde V$ and a surjective map $\pi:\tilde V\to V$. A MAXCUT instance $\tilde w$ on $\tilde V$ is defined as follows:
$$\tilde w(\tilde x,\tilde y)=\frac{w(x,y)}{|\pi ^{-1}(x)|\cdot |\pi ^{-1}(y)|}$$ where $\pi (\tilde x)=x,\pi (\tilde y)=y$.
It is not hard to prove that
\begin{proposition}
Consider the following map from cuts of $w$ to cuts of $\tilde w$ defined by
$$(S,\bar S)\mapsto(\pi ^{-1}(S),\pi^{-1}(\bar S))$$
Then
\begin{enumerate}
\item This map preserves weights, stability and local stability of cuts.
\item Restricted to the locally stable cuts (i.e., $\gamma$-locally stable cuts with $\gamma>1$), this is a bijection {\em onto} the locally stable cuts of $\tilde w$.
\item It maps maximal cuts to maximal cuts.
\end{enumerate}
\end{proposition}
As the following proposition shows, the above construction is a reduction from metric to $(4+o(1))$-dense instances.
\begin{proposition}
Let $w:V\times V\to [0,\infty)$ be an instance of Metric-MAXCUT with $w(V,V)=2\cdot |V|^2$. Consider the map $\pi:\coprod_{x\in V}[\lfloor \tau_w(x)\rfloor]\to V$. The instance $\tilde w$ obtained by $\pi$ is $(4+o(1))$-dense.
\end{proposition}
\proof
Let $\tilde x,\tilde y\in \tilde V$ such that $\pi (\tilde x)=x,\pi (\tilde y)=y$. It is easy to see that (see \cite{VK})
$2\cdot |V|\cdot \tau_w(x)\ge w(V,V)$, $\lfloor \tau_w(x)\rfloor\ge \left(1-\frac{1}{|V|}\right)\tau_w(x)$, $ \tau_{\tilde w}(\tilde x)=\frac{\tau_w(x)}{\lfloor \tau_w(x)\rfloor}\ge 1$ and $w(x,y) \leq \frac {1}{|V|} (\tau_w(x) +  \tau_w(y))$. Thus, we have
\begin{eqnarray*}
\tilde w(\tilde x,\tilde y) &=&\frac{w(x,y)}{\lfloor \tau_w(x)\rfloor\cdot \lfloor \tau_w(y)\rfloor}\\
&\le& \frac{1}{\left(1-1/|V|\right)^2}\cdot \frac{w(x,y)}{ \tau_w(x)\cdot \tau_w(y)}\\
&\le& \frac{1}{\left(1-1/|V|\right)^2}\cdot \frac{\frac{1}{|V|}[\tau_w(x)+\tau_w(y)]}{ \tau_w(x)\cdot \tau_w(y)}\\
&=&\frac{1}{\left(1-1/|V|\right)^2}\cdot \left(\frac{1}{|V|\tau_w(x)}+\frac{1}{|V|\tau_w(y)}\right)\\
&\le& \frac{1}{\left(1-1/|V|\right)^2}\cdot\frac{4}{w(V,V)}\\
&\le&\frac{1}{\left(1-1/|V|\right)^2}\cdot\frac{4}{|\tilde V|}\\
&=& (4+o(1))\frac{\tau_{\tilde w}(\tilde x)}{|\tilde V|}
\end{eqnarray*}
\proofbox
\begin{corollary}\label{cor:metric}
Let $\epsilon>0$.
\begin{enumerate}
\item  There is a randomized polynomial time algorithm for $(1+\epsilon)$-locally stable instances of Metric-MAXCUT.
\item The number of $(1+\epsilon)$-locally stable cuts in a metric instance is polynomial in $|V|$.
\end{enumerate}
\end{corollary}

\subsubsection{A faster algorithm for $(3+\epsilon)$-stable metric instances}
\begin{proposition}\label{prop:cut-edges-are-large}
Let $(L,R)$ be a $\gamma$-locally stable cut of an instance, $w$, of Metric-MAXCUT. Then, for every $x\in L,z\in R$,
$w(x,z)\ge \left(\frac{\gamma^2-1}{\gamma}\right)\cdot\frac{w(x,R)}{\gamma\cdot|R|+|L|}$.
\end{proposition}
\proof
Using $\gamma$-local stability and the triangle inequality we obtain
\begin{eqnarray*}
\frac{1}{\gamma}w(x,R)\ge w(x,L) &=& \sum_{y\in L}w(x,y)\\
&\ge& \sum_{y\in L}(w(z,y)-w(x,z))\\
&=& w(z,L)-|L|w(x,z)\\
&\ge& \gamma w(z,R)-|L|w(x,z)\\
&=&\gamma \sum_{y\in R}w(z,y)-|L|w(x,z)\\
&\ge& \gamma \sum_{y\in R}(w(y,x)-w(z,x))-|L|w(x,z)\\
&=&\gamma w(x,R)-\gamma |R|w(x,z)-|L|w(x,z)
\end{eqnarray*}
\proofbox

\begin{theorem}
\label{it_is_a_ball}
Let $(X,w)$ be an instance of Metric-MAXCUT and let $(L,R)$ be a $\gamma=(3+\epsilon)$-locally stable cut with $\epsilon > 0$. Then either $L$ or $R$ is a (metric) ball.
\end{theorem}
\proof
W.l.o.g., $|L|\ge \frac{n}{2}$. We find some $x\in L$ such that $\forall z\in R,\; w(z,x)>\operatorname{diam}(L)$, thus proving our claim.
Select some $x,y\in L$ with $w(x,y)=\operatorname{diam}(L)$. For every $z\in L$, we write
$w(x,y)\le w(x,z)+w(y,z).$
Summing over every $z\in L$, this yields
$|L|\cdot w(x,y)\le w(x,L)+w(y,L)$.
W.l.o.g., assume that $w(x,L)\ge \frac{|L|}{2}\cdot w(x,y)$. By local stability,
\begin{equation}\label{eq:1}
w(x,y)\le \frac{2}{|L|}w(x,L)\le \frac{2\cdot w(x,R)}{\gamma\cdot|L|}
\end{equation}
By proposition \ref{prop:cut-edges-are-large}, every $z \in R$ satisfies
$w(x,z)\ge  \left(\frac{\gamma^2-1}{\gamma}\right)\cdot\frac{w(x,R)}{\gamma\cdot|R|+|L|}$.
Combined with equation (\ref{eq:1}), and the assumptions that $\gamma>3$ and $|L|\ge |R|$, we obtain that
$w(x,z)>w(x,y)$
as claimed.
\proofbox
By Theorem \ref{it_is_a_ball}, the maximal cut of $(3+\epsilon)$-locally stable instances of Metric-MAXCUT can be found by simply considering all $O(n^2)$ balls.

\begin{note}
Theorem~\ref{it_is_a_ball} is tight in the following sense. We show an example of  $(3-\epsilon)$-stable metric instance (not just locally-stable), where neither side of its maximal cut is a ball, nor can it even be expressed as the union of few balls.

Here is the example: It is a metric space $(X,w)=(L\coprod R,w)$ where $L=\{l_1,\ldots,l_{2n}\}$, $R=\{r_1,\ldots,r_{2n}\}$. Generally speaking, the distance between two points which are both in $L$ or in $R$ is $1$. The distance between a point in $L$ and a point in $R$ is $3$, the following are exceptions to the general rule:
$\forall 1\le i \le n,\;w(l_{2i-1},l_{2i})=w(r_{2i-1},r_{2i})=2$
and
$\forall 1\le i\le 2n,\; w(l_i,r_i)=2$
It is not hard to see that $w$ is a $(3-o(1))$-stable metric instance and each side of its maximal cut cannot be decomposed into fewer than $2n$ balls.
\end{note}

\section{Distinguished and Expanding Instances}\label{sec:distinct}
Let $w:V\times V\to[0,\infty)$ be an instance of MAXCUT with a maximal cut $(S,\bar S)$. We identify $w$ with an $n\times n$ matrix $W$, where $W_{ij}=w(i,j)$. Define $w_{cut}:V\times V\to\mathbb R$ by $w_{cut}(u,v)=w(u,v)$ for $uv\in E(S,\bar S)$ and $w_{cut}(u,v)=0$ otherwise. Similarly, denote $w_{uncut}=w-w_{cut}$. Denote by $W_{cut}$ and $W_{uncut}$ the matrices corresponding to $w_{cut}$ and $w_{uncut}$ respectively. Finally, let $D^{cut},D^{uncut},D$ and $D'$ be the diagonal matrices defined by $D^{cut}_{ii}=\sum_j W^{cut}_{ij}$, $D^{uncut}_{ii}=\sum_j W^{uncut}_{ij}$, $D=D^{cut}+D^{uncut}$ and $D'=D^{cut}-D^{uncut}$.

\begin{lemma}
If $w$ is $\gamma$-locally stable where $\gamma > \frac{2}{1-\sqrt{1-\left(h(w_{cut})\right)^2}}$, then $W+D'$ is a PSD matrix of rank $n-1$.
\end{lemma}
As shown in \cite{BL} there is an efficient algorithm that correctly solves all instances that satisfy the conclusion of the Lemma (As pointed out in the Appendix, the GW-algorithm solves all such instances). This proves the second part of Theorem \ref{thm:main_distinct}.
\proof First, we note that it is enough to prove that $D^{-\frac{1}{2}}(W+D')D^{-\frac{1}{2}}$ is a PSD matrix of rank $n-1$. 
Let $f:V\to\mathbb R$ be the vector defined by $f_i=\sqrt{D_{ii}}$ for $i\in S$ and $f_i=-\sqrt{D_{ii}}$ for $i\in \bar S$. Since $f^TD^{-\frac{1}{2}}(W+D')D^{-\frac{1}{2}}f=0$, it is enough to show that $v^TD^{-\frac{1}{2}}(W+D')D^{-\frac{1}{2}}v> 0$ for every unit vector $v$ that is orthogonal to $f$. Note that
\begin{equation}\label{eq:4}
D^{-\frac{1}{2}}(W+D')D^{-\frac{1}{2}}=D^{-\frac{1}{2}}(D^{cut}+W^{cut}-D^{uncut}+W^{uncut})D^{-\frac{1}{2}}
\end{equation}
The matrix $D^{-\frac{1}{2}}(W^{cut}+D^{cut})D^{-\frac{1}{2}}$ is positive semi-definite and $f$ is in its kernel (to see that, note that for $u\in\mathbb R^n$, $u^T(W^{cut}+D^{cut})u=\sum_{ij}W^{cut}_{ij}(u_i+u_j)^2$). Therefore we have
\begin{equation}\label{eq:5}
v^T D^{-\frac{1}{2}}(W^{cut}+D^{cut})D^{-\frac{1}{2}}v\ge \lambda_2
\end{equation}
where $0=\lambda_1\le \lambda_2\le\ldots\le \lambda_n$ are the eigenvalues of $D^{-\frac{1}{2}}(W^{cut}+D^{cut})D^{-\frac{1}{2}}$. Moreover, $W^{uncut}+D^{uncut}\succeq 0\Rightarrow 2D^{uncut}\succeq D^{uncut}-W^{uncut}$, where $A\succeq B$ means that the matrix $A-B$ is PSD.  Thus, we have,
\begin{equation}\label{eq:6}
v^T D^{-\frac{1}{2}}(D^{uncut}-W^{uncut})D^{-\frac{1}{2}}v\le 2\cdot v^TD^{-\frac{1}{2}}D^{uncut}D^{-\frac{1}{2}}v\le 2\cdot \max_{i}\frac{D^{uncut}_{ii}}{D_{ii}}\le \frac{2}{\gamma+1}
\end{equation}
Combining equations (\ref{eq:4}), (\ref{eq:5}) and (\ref{eq:6}), it is enough to show that $\lambda_2 >  \frac{2}{\gamma+1}$. However, since $w_{cut}$ is bipartite, the matrices $D^{-\frac{1}{2}}(D^{cut}+W^{cut})D^{-\frac{1}{2}}$ and $D^{-\frac{1}{2}}(D^{cut}-W^{cut})D^{-\frac{1}{2}}$ have the same spectrum\footnote{To see that, let $P:\mathbb R^n\to\mathbb R^n$ be the operator that multiply by $-1$ the coordinates corresponding to one side of the cut and fixes the other. The operator $P$ commute with diagonal matrices and satisfies $WP=-PW$. Thus, $v$ be an eigenvector of $D^{-\frac{1}{2}}(D^{cut}+W^{cut})D^{-\frac{1}{2}}$ with an eigenvalue $\lambda$ iff $Pv$ an eigenvector of $D^{-\frac{1}{2}}(D^{cut}+W^{cut})D^{-\frac{1}{2}}$ with an eigenvalue $\lambda$.}. Also, $D^{-\frac{1}{2}}(D^{cut}-W^{cut})D^{-\frac{1}{2}}$ and $D^{-1}(D^{cut}-W^{cut})$ have the same spectrum\footnote{Since $v$ is an eigenvector of $D^{-\frac{1}{2}}(D^{cut}-W^{cut})D^{-\frac{1}{2}}$ with eigenvalue $\lambda$ iff  $D^{-\frac{1}{2}}v$ is an eigenvector of $D^{-1}(D^{cut}-W^{cut})$ with eigenvalue $\lambda$.} so it suffices to show that $\mu_2 > \frac{2}{\gamma+1}$, where $\mu_2$ is the second smallest eigenvalue of $D^{-1}(D^{cut}-W^{cut})$. By the known relation between expansion and the second eigenvalue of the Laplacian (e.g., Theorem 2.2 in \cite{FN}), it follows that
$\mu_2\ge \min_{i}\frac{D^{cut}_{ii}}{D_{ii}}\cdot (1-\sqrt{1-h(w_{cut})^2})\ge \frac{\gamma}{\gamma+1}(1-\sqrt{1-h(w_{cut})^2})$
\proofbox
Finally, to prove the first part of Theorem \ref{thm:main_distinct}, it is enough to show that if $w$ is $\alpha$-distinguished then $h(w_{cut})\ge \alpha$. Indeed, for $\emptyset\ne A\subset V$ we have
$$\tau_{w_{cut}}(A)
=\xi_{w}(A)
\ge \xi_{w}(A)-\iota_{w}(A)
\ge \alpha \cdot \min\{\mu_w(A),\mu_w(\bar A)\}
\ge \alpha \cdot \min\{\mu_{w_{cut}}(A),\mu_{w_{cut}}(\bar A)\} $$

\section{Algorithms for stable instances}\label{sec:general}
We begin with a useful observation.
\begin{observation}\label{oservation:mereging_vertices}
Let $w$ be a $\gamma$-stable instance of MAXCUT, and let $w'$ be obtained from $w$ by merging two vertices\footnote{Let $w:V\times V\to \mathbb R$ be an instance and let $v,u \in V$. The instance $w':V'\times V'\to \mathbb R$ obtained upon {\bf merging $v,u$} is defined as follows. $V'=V\setminus\{u,v\}\cup\{v'\}$ and $w'(x,y)=w(x,y)$ for $x,y\in V\setminus \{v,u\}$, also, $w'(v',x)=w(v,x)+w(u,x)$.} on the same side of $w$'s maximal cut. Then $w'$ is $\gamma$-stable and its maximal cut is induced from $w$'s maximal cut.
\end{observation}
By the above observation, we conclude that in order to design an efficient algorithm for $\gamma$-stable instances, it is enough to show in every $\gamma$-stable instance, we can efficiently find a pair of vertices that are on the same side of the cut. Once two such vertices are found, we merge them and proceed recursively. This applies as well when $\gamma$ is not a constant, but a non-decreasing function of $n$.

As an easy warm-up, we show how this observation yields a simple efficient algorithm that solves every $2n$-stable instance $w:V\times V\to\mathbb R$ of MAXCUT. This is a simplification of an algorithm from~\cite{BL}. By observation \ref{oservation:mereging_vertices}, it suffices to find two vertices which are on the same side of the maximal cut. Pick an arbitrary vertex $v\in V$. If $vu$ is the heaviest edge incident with $v$, then clearly $w(v,u)\ge \frac{1}{n-1}\tau(v)$. On the other hand, by observation \ref{observation:equiv_def}, $\iota(v)\le \frac{1}{2n+1}\tau(v)$, so $w(v,u)>\iota(v)$ and we conclude that $vu$ is a cut edge. Now, let $e$ be the heaviest edge incident with $\{u,v\}$, say $e=vz$. Again, $w(v,z)\ge\frac{1}{2(n-2)}\tau(\{u,v\})$ and by observation \ref{observation:equiv_def}, $\iota(\{v,u\})\le \frac{1}{2n+1}\tau(\{v,u\})$, implying that $w(v,z) > \iota(\{v,u\})$. Consequently $vz$ is a cut edge. But since $vz$ and $vu$ are cut edges, the vertices $z$ and $u$ are on the same side of the cut.

\subsection{A deterministic algorithm for $O(\sqrt{n})$-stable instances}
Following observation \ref{oservation:mereging_vertices}, the algorithm we present will find two vertices which are on the same side of the cut. Let $w:V\times V\to \mathbb R$ be a $\gamma$-stable instance of MAXCUT with $\gamma>\sqrt{8n+4}+1$ and let $(S,\bar S)$ be a maximal cut. We first deal with very heavy edges. Define
$$T^1:=\{vu:w(v,u)>\frac{1}{\gamma+1}\mu(v)\}$$
By observation \ref{observation:equiv_def}, all edges in $T^1$ are cut edges. Thus if there are two incident edges $uv,vz\in T^1$, then $u$ and $z$ are on the same side of the cut and we are done. It remains to consider the case where $T^1$ is a matching. Define
$$T^2=\{uv\notin T^1:w(u,v)>\frac{1}{\gamma+1}\tau(\{u,z\})\text{ for some $uz\in T^1$}\}$$
Again, by observation \ref{observation:equiv_def}, all edges in $T^2$ are cut edges. If $T^2$ is nonempty, say $uv\in T^2$, then there exists some $uz\in T^1$ with $w(u,v)>\frac{1}{\gamma+1}\tau(\{u,z\})$, which implies that $v$ and $z$ are on the same side of the cut. We proceed to consider the case where $T^2$ is empty.

For every $u,v\in V$ define
$$\tilde w(u,v)=\begin{cases}
0 & vu\in T^1\\
w(u,v) & o/w
\end{cases},\;\hat w(v)=\begin{cases}
\tau(\{u,v\}) & vu\in T^1\text{ for some $u\in V$}\\
\tau(v) & o/w
\end{cases}$$
Note that $\hat{w}(v)$ is well defined, since $T^1$ is a matching by assumption. Since $T^2=\emptyset$ and $T^1$ is a matching, we have, for every $u\in V$,
$\tilde{w}(v,u)\le\frac{1}{\gamma+1}\hat{w}(v)$
and, again by observation \ref{observation:equiv_def},
$\iota(v)\le\frac{1}{\gamma+1}\hat{w}(v)$.
Next, we observe as well that separated vertices cannot have too many common neighbors. For $u,v\in V$ we define
$n(u,v):=\sum_{z\in V}\tilde w(v,z)\tilde w(z,u)$.
If $v$ and $u$ are separated, say $v\in S,u\in \bar S$, then
\begin{eqnarray*}
n(u,v) &=& \sum_{z\in \bar S}\tilde w(v,z)\tilde w(z,u)+\sum_{z\in S}\tilde w(v,z)\tilde w(z,u)\\
&\le & \frac{1}{\gamma+1}\hat w(v)\cdot \iota(u)+\frac{1}{\gamma+1}\hat w(u)\cdot \iota(v)\\
&\le & \frac{2}{(\gamma+1)^2}\hat w(u)\cdot\hat w(v)
\end{eqnarray*}
Thus, it suffices to find two vertices $v,u$ with $n(u,v)>\frac{2}{(\gamma+1)^2}\hat w(u)\cdot\hat w(v)$, and place them on the same side of the cut. Indeed, if no such pair exists we have

\begin{eqnarray*}
\frac{1}{4}\sum_{v\in V}\hat w^2(v) &\le & \sum_{v\in V}\tau_{\tilde w}^2(v)\\
&=& \sum_{u,v,z\in V}\tilde w(u,z)\tilde w(z,v)\\
&=&\sum_{u,v\in V,\;u\ne v}n(u,v)+\sum_{u,z\in V}\tilde w^2(u,z)\\
&\le&\frac{2}{(\gamma+1)^2}\sum_{u,v\in V,\;u\ne v}\hat w(u)\hat w(v)+\sum_{u\in V}\frac{1}{\gamma+1}\hat w(u)\sum_{z\in V}\tilde w(u,z)\\
&\le & \frac{2}{(\gamma+1)^2}(\sum_{u\in V}\hat w(u))^2+\frac{1}{\gamma+1}\sum_{u\in V}\hat w(u)\tau_{\tilde w}(u)\\
&\le & \frac{2n}{(\gamma+1)^2}\sum_{u\in V}\hat w^2(u)+\frac{1}{\gamma+1}\sum_{u\in V}\hat w^2(u)
\end{eqnarray*}
And it follows that $\gamma\le \sqrt{8n+4}+1$. A contradiction.

\section{Conclusion and open problems}
Our results together with work from \cite{AB,ABS,BL,DLS,BL2} show that in many cases practically interesting instances of hard problems are computationally feasible. Still much remains to be done toward a new paradigm of analyzing the complexity of computational problems of practical significance. Even if we restrict our attention to MAXCUT, many problems remain open. Here are some of the more significant challenges:

\begin{itemize}
\item
Following \cite{BL}, we recall the (admittedly bold) conjecture that there is a constant $\gamma^*>1$, s.t. $\gamma^*$-stable instances can be solved in polynomial time.
\item It is interesting seek the best possible dependency of $\gamma$ on $\alpha$ in Theorem~\ref{thm:main_distinct}. We are quite certain that further improvements are possible.
\item With reference to Corollary~\ref{cor:metric}, can you find a {\em practically} efficient algorithm for, say, $2$-locally stable metric instances?
\end{itemize}

\appendix
\section{The Spectral approach and the GW algorithm}\label{sec:spectral}
Convex programming relaxations play a key role in the study of hard computational problem. They mostly play a prominent role in the search for {\em approximate} solutions. The Goemans-Williamson (GW) approximate solution for MAXCUT is a prime example of this approach. Can such algorithms provide as well {\em exact} solutions for {\em practically interesting} instances? Many papers (e.g. \cite{B,DP,GM,M}) study the relationships between the maximal cut and spectrum of matrices associated with the instance. Such ideas have led to various heuristics and approximation algorithms for MAXCUT. In section \ref{sec:spectral} we ask under which conditions those methods solve MAXCUT exactly. As shown is Section \ref{sec:distinct}, distinguished instances satisfy such conditions.

We need some terminology. We identify an instance $w$ of MAXCUT with an $n\times n$ matrix $W$, where $W_{ij}=w(i,j)$. A vector $v\in \mathbb{R}^n$ is called a {\em generalized least eigenvector (GLEV)} of $W$ if there is a diagonal matrix $D$ such that $v$ it is an eigenvector of $W+D$, corresponding to $(W+D)$'s least eigenvalue, $\lambda$. By letting $\Delta:=D-\lambda I$ we see that $v$ is a GLEV iff $v$ is in the kernel of $W+\Delta$ for $\Delta$ diagonal with $W+D\succeq 0$. (As usual $A\succeq 0$ means that $A$ is positive semi-definite). A vector $v\in\mathbb R^n$ {\em induces} the cut $(S,\bar S)$ where $S=\{i:v_i>0\}$. An algorithm for MAXCUT is called {\em spectral} if it always returns a cut that is induced by a GLEV.  

Many popular approximation algorithms and heuristics for MAXCUT are spectral. They usually work by returning the cut induced by $w$'s lowest eigenvector (LEV) or by LEV's of related matrices. As we note below, the GW-algorithm is also spectral. Here is the underlying logic of this approach. The {\em characteristic vector} of the cut $(S,\bar S)$ is defined as $\delta_S=\chi_S-\chi_{\bar S}$ where $\chi_A:V\to\{0,1\}$ is the indicator function of $A$. If $D$ is a diagonal matrix, then $\delta_S^T(W+D)\delta_S=2w(V)+\sum_{i=1}^n(D_{ii}-W_{ii})-4w(S,\bar S)$. Thus, the MAXCUT problem can be formulated as follows
\begin{equation}\label{eq:maxcut_as_binary_opt}
\begin{aligned}
& \text{minimize}& &v^T(W+D)v\\
& \text{subject to} & & v\in \{1,-1\}^n
\end{aligned}
\end{equation}
A natural relaxations to this problem is.
\begin{equation}\label{eq:maxcut_relaxation}
\begin{aligned}
& \text{minimize}& &v^T(W+D)v\\
& \text{subject to} & & ||v||=1
\end{aligned}
\end{equation}
where $||\cdot||$ denotes the Euclidean norm.
Now the set of solutions $v$ of (\ref{eq:maxcut_relaxation}) coincides with the set of least eigenvectors of $W+D$. In view of (\ref{eq:maxcut_as_binary_opt}), it is natural to consider the cut induced by such $v$.

\subsubsection*{The GW-Algorithm}
There is another relaxation to (\ref{eq:maxcut_as_binary_opt}), that seems unrelated to (\ref{eq:maxcut_relaxation}). It was suggested by \cite{GM} and will play a major role in the sequel. In problem (\ref{eq:maxcut_as_binary_opt}) we seek $n$ vectors $v_1,\ldots, v_n$ in the $0$ dimensional sphere $S^0=\{-1,1\}$ to minimize $\sum_{i,j}W_{i,j}\langle v_i,v_j\rangle$. Interesting relaxations are obtained by replacing $S^0$ with $S^m$ for some $m$. As observed by \cite{GM} for $m=n-1$, the relaxation
\begin{equation}\label{eq:GW_vectors}
\begin{aligned}
& \text{minimize}& & \sum_{i,j}W_{i,j}\langle v_i,v_j\rangle \\
& \text{subject to} & & v_i\in S^{n-1}
\end{aligned}
\end{equation}
is feasible. In the ideal case, the solution $v_1,\ldots ,v_n$ of (\ref{eq:GW_vectors}) is contained in a copy of $S^0$, embedded in $S^{n-1}$. which makes it a solution for (\ref{eq:maxcut_as_binary_opt}) (in its new formulation). Thus, in the ideal case, separated vectors correspond to two antipodal points, and all vertices that are on the same side of the cut get mapped to the same point. Even if this ideal picture does not hold, one may expect that the angle between separated vertices be large. Therefore, to extract a cut from $v_1,\ldots,v_n$ we need a method that tends to (combinatorially) separate vertices whose images on the sphere are far apart. In \cite{GM} this is done by returning the cut induced by the vector $u\in\mathbb{R}^n$ defined by $u_i=\langle v,v_i\rangle$ where $v\in S^{n-1}$ is sampled uniformly. This yields the approximation ratio $0.879$.

To solve (\ref{eq:GW_vectors}) the GW algorithm finds first a solution $P$ to the problem
\begin{equation}\label{eq:GW_primal}
\begin{aligned}
& \text{minimize}& &P\circ W\\
& \text{subject to} & & P\succeq 0\\
& & &  P_{ii}=1,\;\forall i\in [n]
\end{aligned}
\end{equation}
Where $P\circ W:=\sum_{1\le i,j\le n}P_{ij}\cdot W_{ij}$. Since $P\succeq 0$
it is possible to find next vectors $v_1,\dots, v_n$ such that $P_{ij}=\langle v_i,v_j\rangle$. The dual to (\ref{eq:GW_primal}) is (see \cite{GM})
\begin{equation}\label{eq:gw_dual}
\begin{aligned}
& {\text{maximize}}& &\sum_{i=1}^n D_{ii}\\
& \text{subject to} &  & W-D \succeq 0.\\
& & & D \text{ is diagonal}
\end{aligned}
\end{equation}

As observed in \cite{GM}, by SDP duality the optima of (\ref{eq:GW_primal}) and (\ref{eq:gw_dual}) coincide. Denote by $\cp(W)$ and $\cd(W)$ the set of optimal solutions to (\ref{eq:GW_primal}) and (\ref{eq:gw_dual}) respectively. Denote also $\cp=\{P\in M_n(\mathbb R):P\succeq 0 \text{ and } \forall i,\;P_{ii}=1\}$, $\cd=\{D\in M_n(\mathbb R):D\text{ is diagonal}\}$. We say that $W$ is {\em GW-bipolar} if there exists a solution to (\ref{eq:gw_dual}) that also solves the binary problem (\ref{eq:maxcut_relaxation}) (i.e., it is contained in a copy of $S^0$ embedded in $S^{n-1}$). Equivalently, $W$ is GW-bipolar if $\cal P \rm(W)$ contains a matrix of the form $v\cdot v^T$ for some $v\in \{-1,1\}^n$. Finally, we shall say that $W$ is {\em strongly GW-bipolar} if {\em every} solution to (\ref{eq:GW_vectors}) is also a solution of (\ref{eq:maxcut_relaxation}). Our interest in strongly GW-bipolar instances is clear. The maximal cut of such an instance can be immediately read of the output of the GW-algorithm.

{\bf An overview.}
We start by asking which instances of MAXCUT can be solved {\em exactly} by a spectral algorithm. As we show, the maximal cut is induced by a $\pm 1$ GLEV iff the instance is GW-bipolar. More generally, an instance can be correctly solved by {\em some} spectral algorithm iff it is has a certain perturbation that is GW-bipolar. This provides additional motivation to the study of GW-bipolar instances.

We give a primal-dual characterizing of the set of solutions to the GW-relaxation. Specifically, we show that the dual GW problem (\ref{eq:gw_dual}) always has a unique solution $D$ and the solutions of the primal problem are $\mathcal{P}(W)=\{ P\in\mathcal{P}:P\cdot(W-D)=0 \}$.
This allows us to conclude that the GW-algorithm is a spectral algorithm according to our definition. We also show that GW-bipolarity is equivalent to a condition from \cite{BL}, under which MAXCUT can be solved exactly in polynomial time.

\subsection{Cuts induced by GLEV's}\label{sec:induced-by-levs}
Let $w:V\times V\to \mathbb R^+$ be an instance with an associated matrix $W$. We seek conditions under which a given cut $S$ is induced by GLEV. Let  $v\in \reals^V$ be a vector that induces the cut $S$. As noted before, $v$ is a GLEV if and only if $v$ is in the kernel of $W+D$ for some diagonal matrix $D$ for which $W+D\succeq 0$. Thus, $v$ is a GLEV of $W$ if and only if the optimum of the following SDP is $0$.
\begin{equation}\label{eq:induced_by_LEV_primal}
\begin{aligned}
& \underset{P}{\text{minimize}}& &v^T(W+D)v\\
& \text{subject to}
& & W+D	 \succeq 0\\
& & & D\text{ is diagonal}
\end{aligned}
\end{equation}

The dual program of (\ref{eq:induced_by_LEV_primal}) is
\begin{equation}\label{eq:induced_by_LEV_dual}
\begin{aligned}
& \underset{P}{\text{maximize}}& &v^TWv-P\circ W\\
& \text{subject to}
& & P_{ii}=v_i^2\\
& & & P	 \succeq 0
\end{aligned}
\end{equation}
Since (\ref{eq:induced_by_LEV_primal}) has a positive definite solution, strong duality holds. Thus, $v$ is a GLEV iff the optimum of (\ref{eq:induced_by_LEV_dual}) is $0$.

Now, the optimum of the dual is $0$ iff the perturbation of $W$ defined by $W'_{ij}=|v_i|\cdot |v_j|\cdot W_{ij}$ is GW-Bipolar. To see that, note that the mapping $P'\mapsto P$ where $P_{ij}=|v_i|\cdot |v_j|\cdot P'_{ij}$ maps the feasible solutions to the primal GW-relaxation (\ref{eq:GW_primal}) for $W'$ onto the feasible solution to (\ref{eq:induced_by_LEV_dual}). Moreover, $P\circ W=P'\circ W'$. Thus, the optimum of (\ref{eq:induced_by_LEV_dual}) is zero iff the optimum of the primal GW relaxation of $W'$ is $v^TWv=\delta_S^TW'\delta_S$. Consequently, the optimum of (\ref{eq:induced_by_LEV_dual}) is $0$ iff the optimum of (\ref{eq:GW_primal}) is attained by a $\pm 1$ vector, making $W'$ GW-bipolar. Note that if $v$ "strongly induces" the cut $S$ -- that is, if all coordinates $|v_i|$ are roughly equal, then $W'$ is just a small perturbation of $W$. Taking this to the extreme, we conclude that the cut is induced by a $\pm 1$ GLEV iff $W$ is GW-bipolar.

\subsection{The GW algorithm and GW-bipolar instances}\label{sec:gw_char}
We start with a primal-dual characterization of $\cd(W)$ and $\cp(W)$.

\begin{theorem}\label{thm:gw_primal_dual_char}
Let $W$ be a non-negative symmetric matrix with $0$-diagonal. Then,
\begin{enumerate}
\item $\cd(W)$ is a singleton\footnote{Henceforth we usually do not distinguish between $\cd(W)$ and the single matrix that it contains.}.\label{part_1}
\item $\cp(W)=\{P\in \cp:P(W-\cd(W))=0\}$\label{part_2}
\end{enumerate}
\end{theorem}
\begin{lemma}\label{lemma:gw_primal_dual}
For every $D^0\in\cd(W),\;P^0\in \cp(W)$ we have
$$\cp(W)=\{P\in \cp:P(W-D^0)=0\}$$
$$\cd(W)=\{D\in \cd:(W-D)\succeq 0,\;P^0(W-D)=0\}$$
\end{lemma}
\proof
Let $D^0\in \cd(W)$, $P\in \cp$. By strong duality,
$$P\in \cp(W)\Leftrightarrow W\circ P=\sum_{i=1}^nD_i^0 \Leftrightarrow W\circ P=D^0\circ P $$
Since $W-D^0$ and $P$ are PSDs, $P\circ(W-D^0)=0\Leftrightarrow P(W-D^0)=0$. Thus,
$$\cp(W)=\{P\in \cp:P(W-D^0)=0\}$$
Similarly, let $P^0\in \cp(W)$, $D\in \cd$ such that $W-D\succeq 0$ then
$$D\in \cd(W)\Leftrightarrow W\circ P^0=\sum_{i=1}^nD_i \Leftrightarrow W\circ P^0=D\circ P^0$$
Thus
$$\cd(W)=\{D\in \cd:(W-D)\succeq 0,\;P^0(W-D)=0\}$$
\proofbox

\proof (of Theorem \ref{thm:gw_primal_dual_char})
Part \ref{part_2} follows from part \ref{part_1} and Lemma \ref{lemma:gw_primal_dual}, so it only remains to prove part \ref{part_1}.
Fix some $P^0\in \cp(W)$ and let $D\in \cd(W)$. By considering the $(j,j)$ entry of $P^0(W-D)=0$, we have
$$D_{jj}=\sum_{i=1}^nP^0_{ji}W_{ij}$$
which determines $D$ uniquely.
\proofbox

\begin{corollary}\label{cor:GW_is_spectral_alg}
GW is a spectral algorithm.
\end{corollary}
\proof
Suppose that the optimum of the GW-relaxation is attained at $P$ and let $v_1,\ldots,v_n\in S^{n-1}$ be vectors such that $P_{ij}=\langle v_i, v_j\rangle$. Let $v\in S^{n-1}$ be the vector sampled by the algorithm and let $\sum_{j=1}^n\alpha_jv_j$ be its orthogonal projection on $span \{v_1,\ldots,v_n\}$. The cut returned by the algorithm is the one induced by the vector
$u_i=\langle v, v_i\rangle=\sum_j\alpha_jP_{ij}$. The vector $u$ is a linear combination of $P$'s columns. Thus, by Theorem \ref{thm:gw_primal_dual_char} it is in the kernel of the PSD matrix $W-\cd(W)$.
\proofbox

\begin{corollary}\label{cor:gw_solves_very_stable_instances}
The GW algorithm correctly solves $\Omega(n^3)$-stable instances.
\end{corollary}
\proof
In \cite{BL} it is shown that if $u$ is a GLEV of a $\gamma$-stable instance $W$ such that $\gamma\ge \frac{\max_{(i,j)\in E}|u_iu_j|}{\min_{(i,j)\in E}|u_iu_j|}$ then $u$ induces the optimal cut.
Let $u$ be defined as in the proof of Corollary \ref{cor:GW_is_spectral_alg}. As shown, $u$ is a GLEV. Moreover, by an easy probabilistic argument, w.h.p., $\forall j,n^{-1.5}\le |u_j|\le 1$.
\proofbox
Here is a characterization of GW-bipolar matrices.

\begin{theorem}\label{thm:gw_bipolar_characterization}
Let $W$ be an instance for MAXCUT with maximal cut $S$. Denote $v=\delta_S$ and let $D$ be the diagonal matrix defined by $D_{ii}=-v_i\sum_j{W_{ij}v_j}$. The following conditions are equivalent.
\begin{enumerate}
\item $W$ is GW-bipolar.\label{equiv_1}
\item $\delta_S$ is a GLEV of $W$.\label{equiv_2}
\item $W+D\succeq 0$ \label{equiv_3}
\item The optimum of the dual of the GW-relaxation is attained at $-D$.\label{equiv_4}
\end{enumerate}
\end{theorem}
\proof
As shown in section \ref{sec:induced-by-levs} condition \ref{equiv_1} is equivalent to condition \ref{equiv_2}. Suppose now that \ref{equiv_3} holds. It is not hard to see that $\delta_S$ is in the kernel of $W+D$, so \ref{equiv_2} holds. Condition \ref{equiv_4} clearly entails condition \ref{equiv_3}. Finally, suppose that \ref{equiv_1} holds. Let $D'$ be the solution of problem (\ref{eq:gw_dual}). Since $W$ is GW-bipolar, $\delta_S\cdot \delta_S^T$ is an optimal primal solution. By Lemma \ref{lemma:gw_primal_dual} we deduce that $\delta_S\in ker(W-D')$. It follows that $D'=-D$ and \ref{equiv_4} holds.
\proofbox
As noted before, strongly GW-bipolar instances can be efficiently solved using the GW algorithm. In fact, for those instances there is no need to choose a random vector to produce a cut. Moreover, those instances can be solved simply by taking the sign pattern of the least eigenvector of $W+D$ where $D$ is the solution to problem (\ref{eq:gw_dual}). As we explain next, strong GW-bipolarity is just slightly stronger than GW-bipolarity. Let $W$ be a GW-bipolar instance with maximal cut $(S,\bar S)$. Let $W'$ be the $(1+\epsilon)$-perturbation of $W$ that is obtained by multiplying cut edges by $1+\epsilon$ with $\epsilon > 0$ arbitrarily small. We claim that it is strongly GW-bipolar.
Let $D$ be the diagonal matrix defined in Theorem \ref{thm:gw_bipolar_characterization}. We have $W+D\succeq 0$ if and only if for every $u\in S^{n-1}$
\begin{equation}\label{eq:2}
u^T(W+D)u=\sum_{ij\in E(S,\bar S)}W_{ij}(u_i+u_j)^2
-\sum_{ij \notin E(S,\bar S)}W_{ij}(u_i-u_j)^2\ge 0
\end{equation}
Inequality (\ref{eq:2}) clearly holds for $W'$ as well making it GW-bipolar. Moreover, since the maximal cut is connected, if $u\ne \pm \frac{1}{\sqrt{n}}\cdot\delta_S$ then $\sum_{ij\in E(S,\bar S)}W'_{ij}(u_i+u_j)^2>\sum_{ij\in E(S,\bar S)}W_{ij}(u_i+u_j)^2$. Thus, $u^T(W'+D')u>u^T(W+D)u\ge 0$ where $D'$ is the matrix corresponding to $W'$ from Theorem \ref{thm:gw_bipolar_characterization}. Thus, the matrix $W'+D'$ has rank $n-1$. By Theorem \ref{thm:gw_primal_dual_char} we conclude that $\delta_S\cdot \delta_S^T$ is the only solution to the primal GW-problem for $W'$, making $W'$ strongly GW-bipolar.

\section{A randomized algorithm for $\epsilon\cdot\frac{n}{\log(n)}$-stable instances}
We now describe a simple randomized algorithm that correctly solves $\epsilon\cdot\frac{n}{\log(n)}$-stable instances of MAXCUT. So let $w:V\times V\to[0,\infty)$ be a $\gamma$-stable instance with $\gamma=\epsilon\cdot\frac{n}{\log(n)}$. Our algorithm proceeds as follows.
\begin{enumerate}
\item Set $V_0=\{v_0\}$ for some $v_0\in V$ and set $E_0=\emptyset$.
\item For $t=1\text{ to } |V|-1$
\begin{itemize}
\item Sample a random edge $v_t u_t \in E(V_{t-1},\bar V_{t-1})$, where the probability of every edge is proportional to its weight.
\item Set $V_t=V_{t-1}\cup\{v_t,u_t\},\;E_t=E_{t-1}\cup\{v_tu_t\}$
\end{itemize}
\item Note that $(V_t,E_t)$ is a tree for every $t$ and for $t=|V|-1$ this is a spanning tree. Return the bipartition corresponding to the two-coloring of this tree.
\end{enumerate}

{\bf Analysis:}
In order to return the maximal cut, it is sufficient (in fact, also necessary) that for every $t$, the edge $v_tu_t$ be in the maximal cut. But, by observation \ref{observation:equiv_def}, the edges in $E(V_t,\bar V_t)$ that are in the maximal cut constitute $\ge \frac{\gamma}{\gamma+1}$ fraction of all the edges in $E(V_t,\bar V_t)$. Thus a lower bound on the success probability of the algorithm can be derived as follows:
\begin{eqnarray*}
\left(\frac{\gamma}{\gamma+1}\right)^{n-1} &\ge& \left(1-\frac{1}{\gamma+1}\right)^n\\
&=&\left(1-\frac{1}{\epsilon\cdot\frac{n}{\log (n)}+1}\right)^n\\
&\ge&\left(1-\frac{1}{\epsilon\cdot\frac{n}{\log (n)}}\right)^n\\
&=& \left(e^{-\epsilon}+o(1)\right)^{\ln (n)}\\
&=& n^{\ln (e^{-\epsilon}+o(1))}=n^{-\epsilon +o(1)}
\end{eqnarray*}
In particular, for $\epsilon$ fixed the process succeeds with probability that is at least inverse polynomial in $n$.

\end{document}